\documentclass{PoS}

\newcommand{\beqn}{\begin{eqnarray}}
\newcommand{\eeqn}{\end{eqnarray}}
\newcommand{\be}{\begin{equation}}
\newcommand{\ee}{\end{equation}}

\newcommand{\ba}{\begin{array}{c}}
\newcommand{\bat}{\begin{array}{cc}}
\newcommand{\ea}{\end{array}}

\newcommand{\bi}{\begin{itemize}}
\newcommand{\ei}{\end{itemize}}

\newcommand{\rcht}{R$\chi$T}

\newcommand{\Frac}[2]{\frac{\displaystyle #1}{\displaystyle #2}}

\newcommand{\cO}{{\cal O}}

\newcommand{\mF}{\mathcal{F}}

\newcommand{\mL}{\mathcal{L}}

\newcommand{\lsim}{\stackrel{<}{_\sim}}

\newcommand{\bra}{\langle}
\newcommand{\ket}{\rangle}

\ShortTitle{Renormalization group equations in resonance chiral theory:\\
the $\pi\pi$ vector form-factor}

\abstract{
The use of the equations of motion and meson field redefinitions
allows the simplification of the subleading operators required 
in the one-loop resonance chiral theory calculation of the 
$\pi\pi$--vector form-factor.  
The study of the renormalization group equations of the relevant parameters  
shows the existence of an infrared fixed point for all the couplings.  
It is important to remark that this result does not rely on the high-energy 
form-factor constraints, which are often considered in other works.   
The possibility of developing a perturbative $1/N_C$ expansion
in the slow running region around the fixed point is shown here.
}
\FullConference{
Sixth International Workshop on Chiral Dynamics\\ July 6-10 2009,
Bern, Switzerland.
\\
\vspace*{0.5cm}
This proceedings are based on Ref.~\cite{RGE}.
It has been supported in part by the
CICYT-FEDER-FPA2008-01430, SGR2005-00916, SGR2009-894,
the Spanish Consolider-Ingenio 2010 Program CPAN (CSD2007-00042),
the Juan de la Cierva Program  and
the EU Contract No. MRTN-CT-2006-035482,
\lq\lq FLAVIAnet''.
I would like to thank the organizers for all their effort and kindness
during the conference.
}

\title{RGE in resonance chiral theory: \\
the $\pi\pi$ vector form-factor}
\author{Juan Jos\'e Sanz-Cillero\\
Grup de F\'\i sica Te\'orica and IFAE, Univ.
Aut\'onoma de Barcelona, 08193 Barcelona, Spain}

\renewcommand{\logo}
{\parbox{\textwidth}{\begin{flushright} \vspace*{2cm}
\end{flushright}\vspace*{-0.3cm}
\includegraphics{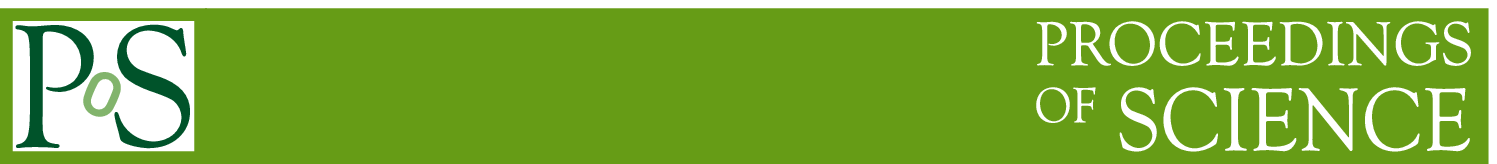}}
}

\begin{document}



\maketitle



Resonance chiral theory (R$\chi$T) is a description  of the Goldstone-resonance
interactions within a chiral invariant framework~\cite{RChTa}.
The pseudo-Goldstone fields $\phi$ are introduce through the exponential  realization
${    u(\phi) =\exp{\left(i \phi/\sqrt{2} F\right)}      }$.  The standard effective field theory
momentum expansion is not valid in the presence of heavy resonance states
and an alternative perturbative counting is required.
R$\chi$T  takes then the formal $1/N_C$ expansion as a guiding principle~\cite{NC}:
at leading order (LO) the interaction terms
in the lagrangian with a number $k$ of meson fields (and their corresponding couplings)
scale as $\sim N_C^{1-\frac{k}{2}}$~\cite{NC}.
At large $N_C$, the resonance fields become classified in $U(n_f)$ multiplets,
with $n_f$ the number of light quark flavours.

The interacting operators of the leading R$\chi$T lagrangian relevant for the
analysis of the $\pi\pi$ vector form-factor (VFF) in the chiral limit are given 
by~\cite{RGE,RChTa,L9}
\begin{eqnarray}
 \mathcal{L}_{\rm LO}^{\mathrm{GB}} \,\,
 =\,\, \Frac{F^2}{4}\, \langle u_\mu u^\mu
 \rangle
 \,,
\qquad
\mathcal{L}_{\rm LO}^V &=\,
 \frac{i\, G_V}{2\sqrt{2}} \bra V_{\mu\nu} [u^\mu, u^\nu] \ket
\, ,  \label{1Rlagrangian}
\end{eqnarray}
The Goldstones fields, given by $u(\phi)$, enter into play through
the covariant tensor
$  u_\mu=i\{
u^\dagger (\partial_\mu - i r_\mu) u- u (\partial_\mu - i\ell_\mu) u^\dagger\}  $.
Likewise, it is convenient to define
$f_\pm^{\mu\nu}=u F_L^{\mu\nu} u^\dagger \pm u^\dagger F_R^{\mu\nu} u$,
with $F_{L,R}^{\mu\nu}$ the left and right field strength 
tensors~\cite{RChTa}.
Here, the antisymmetric tensor field $V^{\mu\nu}$ will be   considered
for the description of the spin--1 mesons~\cite{RChTa,chpt}.

\begin{figure}
\begin{center}
\includegraphics[angle=0,clip,width=5cm]{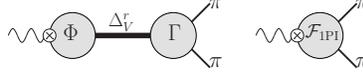}
\caption{{\small 1PI-topologies contributing to the  pion VFF.
 }}
 \label{fig.1PI}
\end{center}
\end{figure}

The VFF can be decomposed in the one-particle irreducible (1PI) topologies of 
Fig.~\ref{fig.1PI}:
\be
\label{eq.NLO-VFF}
\mF(q^2) = \mF(q^2)_{\rm 1 PI} + \Frac{\Phi(q^2) \Gamma(q^2)}{F^2}
\Frac{q^2}{M_V^2 -q^2-\Sigma(q^2) }\ ,
\ee
with $q\equiv p_1+p_2$,
$\Sigma(q^2)$ the vector self-energy, and $\mF(q^2)_{\rm 1 PI}$, $\Phi(q^2)$
and $\Gamma(q^2)$ being provided, respectively,
by the 1PI vertex-functions
for $J_V^\mu\to\pi\pi$, $J_V^\mu\to V$ and ${  V\to\pi\pi  }$.
The one-loop calculation produces a series of ultraviolet divergences
that require of subleading operators in $1/N_C$ ($X_Z$, $X_F$, $X_G$, $\widetilde{L}_9$)
to fulfill the renormalization of the vertex functions~\cite{L9}:
\begin{eqnarray}
\mathcal{L}^{\rm GB}_{\rm NLO} &=&\,
-\, i\, \widetilde{L}_9\bra f_+^{\mu\nu} u_\mu u_\nu\ket \, ,
\nonumber\\
\mathcal{L}^V_{\rm NLO} &=&\,
X_Z\bra V_{\lambda\nu}\nabla^\lambda\nabla_\rho \nabla^2 V^{\rho\nu}\ket
\,+\,  X_F\bra V_{\mu\nu}\nabla^2 f_+^{\mu\nu}\ket
\,+ \,2\, i\, X_G    \bra V_{\mu\nu} \nabla^2 [u^\mu, u^\nu] \ket
\,.
\label{eq.lagrNLO}
\end{eqnarray}
However,   since the  subleading $\mL^V_{\rm NLO}$ operators
are proportional to the equations
of motion, one finds that
$\mathcal{L}^V_{\rm NLO}$ can  be fully transformed
into the $M_V$, $F_V$, $G_V$ and $\widetilde{L}_9$ terms  and
into other operators that  do not contribute
to the VFF by means of  a convenient meson field redefinition~\cite{RGE,L9}.   

Taking now the one-loop expressions for the 1PI topologies
(simplified after the field redefinitions)~\cite{RGE}
and setting $\mu^2=Q^2$, one gets then the simple form-factor structure~\footnote{
The dilogarithmic
contribution $\Delta_t$ from
the triangle with the $t$--channel vector exchange can be found in Ref.~\cite{RGE} 
and, for the energies we are going to study ($|q^2|\lsim 1$~GeV$^2$),
it has little numerical impact.
},
\begin{eqnarray}
\label{eq.VFF}
&&\mF(q^2) = \,\, -\,\, \Frac{ 2 \,Q^2 \widetilde{L}_9(Q^2)}{F^2}
\,\,\,+\,\,\, \left[1+\Delta_t(q^2)\right]\,\,
\left[1- \Frac{F_V(Q^2) G_V(Q^2)}{F^2}\Frac{Q^2}{M_V^2(Q^2)\,  +\,  Q^2}
\right] \, \, ,
\end{eqnarray}
with  the evolution of the remaining couplings (after the field redefinition) 
with the Euclidean squared momentum $Q^2\equiv -q^2$ prescribed by the
renormalization group equations (RGE)~\cite{RGE},
\begin{eqnarray}
\label{eq.running}
\Frac{1}{M_V^2}\Frac{\partial M_V^2}{\partial\ln\mu^2} &=&  \Frac{n_f}{2}
\Frac{2 G_V^2}{F^2}\Frac{M_V^2}{96\pi^2 F^2}  \, ,
\\
\Frac{\partial G_V}{\partial \ln\mu^2}
&=&  G_V\,\, \Frac{n_f}{2}\Frac{M_V^2}{96\pi^2 F^2} \left( \Frac{ 3 G_V^2}{F^2}-1\right)\, ,
\nonumber\\
\Frac{\partial F_V}{\partial\ln\mu^2} &=& 2 \,G_V\,
\Frac{n_f}{2}\Frac{M_V^2}{96\pi^2 F^2} \left( \Frac{F_V G_V}{F^2}-1\right) \, ,
\nonumber\\
\Frac{\partial \widetilde{L}_9}{\partial\ln\mu^2}
&=&
\Frac{n_f}{2}\Frac{1}{192 \pi^2}\,
\left( \Frac{F_V G_V}{F^2} -1\right)\,
\left(1- \Frac{3 G_V^2}{F^2}\right)\, .
\nonumber
\end{eqnarray}

The  RGE solutions for $M_V$ and $G_V$  form a closed system with the
trajectories given by
\begin{eqnarray}
G_V^2\,\,=\,\, \Frac{F^2}{3}\,\left(\,1\, + \, \kappa^3\, M_V^6 \right)\, ,
\qquad
\Frac{1}{M_V^2} \, +\, \kappa \, f(\kappa M_V^2)\, \, =\, \,
-\,\Frac{2}{3}\, \Frac{n_f}{2}\Frac{1}{96\pi^2 F^2}\,
\ln\Frac{\mu^2}{\Lambda^2}\, ,
\label{eq.solMV}
\end{eqnarray}
with $f(x)= \frac{1}{6}\ln\left(\frac{x^2+ 2 x +1}{x^2 -x +1}\right)
- \frac{1}{\sqrt{3}}\arctan\left(\frac{ 2x-1}{\sqrt{3}}\right)-\frac{\pi}{6\sqrt{3}}
=\cO(x)$, and  $\kappa$ and $\Lambda$  integration constants. Since
$-\frac{2\pi}{3\sqrt{3}}\leq f(x)\leq 0$, the term $\kappa f(\kappa M_V^2)$
in~(\ref{eq.solMV}) becomes negligible for very low momentum,
$\mu\ll \Lambda$, producing a logarithmic running.

\begin{figure}[!t]
\begin{center}
\includegraphics[angle=0,clip,width=6.5cm]{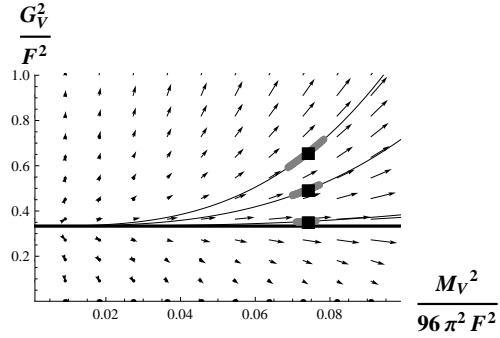}
\caption{{\small
Renormalization group flow for $M_V^2$ and $G_V^2$. The points $M_V(\mu_0)=775$~MeV and
${  G_V(\mu_0)=75,\, 65,\, 55  }$~MeV are plotted with filled squares, together with their
trajectories for ${  n_f=2  }$ (thin black lines).
The horizontal line represents the $G_V$--fixed point at $G_V^2=F^2/3$.
 }}
\label{fig.MVGV}
\end{center}
\end{figure}

The parameters $M_V$ and $G_V$ show then an infrared fixed point at
$M_V=0$ and $G_V=F/\sqrt{3}$.
The corresponding flow diagram  is shown in Fig.~\ref{fig.MVGV}.
The same happens for $F_V$ and $\widetilde{L}_9$,
which freeze out when $\mu\to 0$. $F_V$ tends to the infrared fixed point $\sqrt{3} F$
(and hence $F_V G_V\stackrel{\mu\to 0}{\longrightarrow} F^2$)
and $\widetilde{L}_9(\mu)$ goes to a constant value $\widetilde{L}_9(0)$.
%

Notice that the value of the resonance
couplings at the infrared fixed  point, $F_V G_V=F^2$ and $3 G_V^2=F^2$,
coincides with those obtained if one demands at large--$N_C$ the proper
high energy behaviour of, respectively,
the VFF~\cite{L9,spin1} and the partial-wave scattering amplitude~\cite{scat}.
This interplay between  fixed points  and short-distance behaviour 
is explained with more detail in~\cite{RGE}.
This is also related there with value obtained from
the requirement that our one-loop form factor~(\ref{eq.VFF})
vanishes when $Q^2\to\infty$~\cite{spin1,BrodskyLepage,NLOsatura}.
It leads to the same values:   the constraints
$F_V G_V=F^2$ and $3 G_V^2 =F^2$ are required to freeze out the  
running of $\widetilde{L}_9$ and $F_V G_V$,  killing  
the $Q^2 \ln(Q^2)$ and $Q^0 \ln( Q^2)$ short-distance
behaviour in~(\ref{eq.VFF});  
additionally, $\widetilde{L}_9=0$ is needed  in order to remove
the remaining $\cO(Q^2)$ terms at $Q^2\to\infty$.


Independently of any possible high energy matching~\cite{spin1,NLOsatura},
what becomes clear from the RGE analysis in~\cite{RGE} is the existence
of a region in the RGE space of parameters (around the infrared fixed point at
$\mu\to 0$) where the \rcht\ loops produce small logarithmic corrections.
Thus, although we start with a formal  expansion in $1/N_C$,
the perturbative description for the renormalized  R$\chi$T amplitude  
only makes sense in the momentum range in the proximity of the fixed point. 
In an analogous way, although
the fixed order perturbative QCD cross-section calculations
are formally correct for arbitrary $\mu$  (and independent of it),
perturbation theory can only be applied at high energies.

One of the aims of~\cite{RGE}  was to show how the potentially dangerous higher
power corrections arising at next-to-leading order (NLO)~\cite{L9}
actually correspond to a slow logarithmic
running of the couplings of the LO lagrangian.
We made use of the equations
of motion of the theory and meson field redefinitions  to remove
analytical corrections going like higher powers
of the momenta. This  left in the vertex functions just the problematic 
log  terms $Q^4\ln(Q^2)$,
which were  minimized by means of the convenient choice of scale $\mu^2=Q^2$.   
Their evolution was then controlled  by the RGE~(\ref{eq.running}),  
giving place to a slow logarithmic running for 
$M_V$, $F_V$, $G_V$ and $\widetilde{L}_9$.

These considerations are expected to be relevant for the study of
other QCD matrix elements.  In particular, they may play
an important role in the case of scalar resonances. The
width and radiative corrections are usually  rather sizable in the spin--0 channels.
The possible presence of fixed points and slow--running regions in other
amplitudes (e.g. the pion scalar form-factor)   will be studied in future analyses.

\end{document}